\documentclass[twocolumn,showpacs,preprintnumbers,amsmath,amssymb]{revtex4}

\usepackage{epsfig}
\usepackage{subfigure}
\usepackage{graphicx}
\usepackage{dcolumn}
\usepackage{bm}

\preprint{APS/123-QED}

\begin{document}
\title{First principles electron transport: finite-element
  implementation for nanostructures}

%
\author{P. Havu$^1$, V. Havu$^2$, M. J. Puska$^1$, M. H. Hakala$^1$,
A. S. Foster$^1$, and R. M. Nieminen$^1$}
\affiliation{
1) Laboratory of Physics, Helsinki University of Technology,
P.O. Box 1100, FIN-02015 HUT, Finland }
\affiliation{
2) Institute of Mathematics, Helsinki University of Technology,
P.O. Box 1100, FIN-02015 HUT, Finland}

\date{\today}

\begin{abstract}
We have modeled transport properties of nanostructures using the
Green's function method within the framework of the density-functional
theory. The scheme is computationally demanding so that numerical
methods have to be chosen carefully. A typical solution to the
numerical burden is to use a special basis-function set, which is
tailored to the problem in question, for example, the atomic orbital
basis.  In this paper we present our solution to the problem. We have
used the finite element method (FEM) with a hierarchical high-order
polynomial basis, the so-called {\it p}-elements. This method allows the
discretation error to be controlled in a systematic way. The
{\it p}-elements work so efficiently that they can be used to solve
interesting nanosystems described by non-local pseudopotentials.

We demonstrate the potential of the implementation with two different
systems. As a test system a simple Na-atom chain between two leads is
modeled and the results are compared with several previous
calculations. Secondly, we consider a thin hafnium dioxide (HfO$_2$)
layer on a silicon surface as a model for a gate structure of the next
generation of microelectronics.

\end{abstract}


\maketitle


\section{Introduction}

Using small nano-scale lithographic structures, atomic aggregates and
even single molecules, it is possible to fabricate new kind of
electronic devices \cite{gre_datta}. The function and scale of these
devices is based on quantum-mechanical phenomena, and cannot be
described within the classical regime.  Of particular relevance are
the electron transport properties of these nanoscale devices, as this
will determine their effectiveness in, for example,  a new generation
of transistors. As the experimental work on these devices grows,
increasing emphasis is placed on developing a matching theoretical
description \cite{kemian_rev, tausta_rev}. Although some efforts have
included a full description of an electronic circuit \cite{koe_piiri,
teor_piiri}, current research is mainly focused to study single electronic
components.

Density-functional theory (DFT) is widely used in atomistic modeling
of materials properties and recently also properties and phenomena in
nanostructures.  The power of DFT is in its capacity to treat
accurately systems with a hundreds of atoms, yet retain a full
quantum-mechanical treatment.  Although the full justification of use
of the DFT in electron transport calculations is debated
\cite{tausta3, k_pisteet_kristian} we adopt it as a practical scheme
to describe the real systems and devices.

In the Kohn-Sham scheme of DFT the
electron density is calculated using single-particle wave
functions. The explicit use of the wave functions in constructing the
density suffices well in two kinds of systems. Either the system has a
repeating structure so that it can be modeled with periodic boundary
conditions or the system is so small that it can be calculated as a
whole. In nano electronics, however, a system consists usually of a
small finite part, the nanostructure, which is connected to the
surrounding infinite leads. If one enforces periodic boundary
conditions even a large repeating super cell or  calculation volume
can cause finite-size effects with spurious results for electron
transport.


A commonly used solution to this problem, which we have also employed,
is to combine DFT with the Green's function formalism
\cite{gre_datta}. The Green's functions are first constructed for the
semi-infinite leads by using the analytically known or easily
calculated wave functions. Once the Green's function for the combined
nanostructure and leads is constructed, the wave functions are no
longer needed explicitly.  This makes it possible to use open boundary
conditions between the nanostructure and the lead. In this way we have
an effectively infinite system without periodicity, making the
finite-size effects small. It is also possible to calculate the
electric current through the system for a finite bias voltage between
the leads in a self-consistent manner with the electron density. The
ensuing model for the current is analogous to the Landauer-B\"utteker
model \cite{kemian_rev}. We have used non-local pseudopotentials for
modeling atoms, and the ideal metal ``jellium'' model for the leads. The
charge density in the leads can be varied according to the conducting
properties of the leads we wish to model.

The use of Green's functions instead of the explicit use of wave
functions is computationally demanding. This is why special care has
to be taken in choosing the numerical methods. The first
implementations used tight-binding methods
\cite{tight_binding,tight_binding2}, but a more typical solution is to
expand the Green's functions in a special basis tailored for the
system. Common examples are localized atomic orbitals
\cite{transiesta1,transiesta2}, an $O(N)$ optimized basis
\cite{bernholc}, a wavelet basis \cite{wavelet}, full-potential
linearized augmented plane-waves \cite{embedded}, maximally localized
Wannier functions \cite{wannier}, a finite-difference method
\cite{differenssi}, and linear a finite-element method \cite{fem-green}.
Our solution is to use the finite-element method (FEM). It allows a
systematic error control which is especially important in transport
problems as there are many different properties which must be
monitored. For example, the pole of the Green's function can cause
numerical problems. According to our experience electronic tunneling
in particular is sensitive to numerical accuracy.

Besides systematic error control, the FEM has also other good
properties which makes it a natural method for transport problems. It
is a flexible method which allows one to take into account the
geometry of the nano device exactly. Special boundary conditions are
easy to derive without mixing the model with the numerical method and
their implementation is straightforward. Moreover, the local nature of
the basis produces sparse matrices for which efficient solving methods
exist. Varying the size of the elements can be used to reduce the
number of the basis functions and, consequently, the size of the
system as compared to the finite-difference method. This is especially
true for the high-order $p$-method. Finally, there exists a lot of
theoretical work together with tested and reliable tools, such as mesh
generators and optimized linear solvers. These are used as standard
building blocks for any FEM implementation granting easy access to
state-of-the-art algorithms.  Using the FEM new theoretical or
numerical ideas are easy to implement and test.

The structure of the paper is as follows: in Sec. \ref{model} we
describe the model itself in detail, including a discussion of the
formalism of our implementation; in Sec. \ref{examples} we apply the
model to two example systems, a Na atom chain and HfO$_2$-Si interface
between two leads. In Sec. \ref{discussion} we summarize the work.  In
this paper we use atomic units in all equations.


\section{Model}\label{model}

\begin{figure}[htb]
\begin{center}
  \epsfig{file=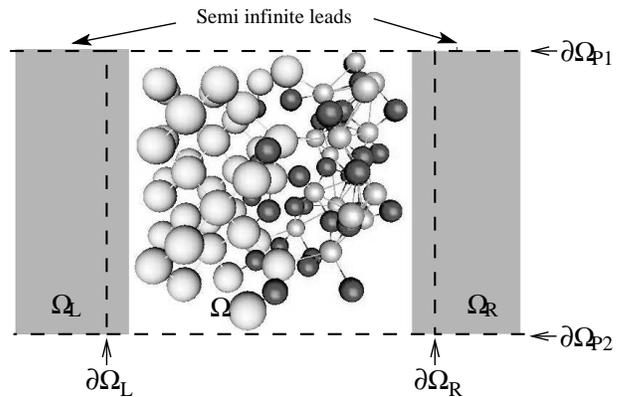,width=0.45\textwidth}
\end{center}
\caption{\label{nano_malli} Schematic picture of the model. The HfO$_2$
  interface is used as an example. The small and large gray gray
  spheres denote the Hf and Si atoms, respectively, and dark spheres the
  O atoms. The gray volumes are the jellium leads. The system consist
  the volumes $\Omega_L$, $\Omega$, and $\Omega_R$ and of the
  boundaries $\partial \Omega_L$, $\partial \Omega_R$, $\partial
  \Omega_{P1}$, $\partial \Omega_{P2}$, $\partial \Omega_{P3}$, and
  $\partial \Omega_{P4}$. }
\end{figure}

The schematic picture of our model is shown in
Fig.~\ref{nano_malli}. Actually, the figure present our second test
case, the HfO$_2$-Si interface between two leads. We have an atomistic
nanostructure between two semi-infinite leads. The system is divided
into three parts, $\Omega$ being the calculation volume, and
$\Omega_L$ and $\Omega_R$ are left and right leads, respectively. The
boundaries $\partial \Omega_{L/R}$ are open so that electrons can
penetrate through them without any reflection or refraction. We use
the DFT to model electron interactions. The basic quantity, the
electron density, is calculated from single-particle Green's
functions. Then we use the density to calculate the effective potential as
%
%
\begin{equation}
V_{\rm eff} = V_{\rm ext} + V_{\rm c} + V_{\rm xc} + V_{\rm bias} +
\hat{V}_{\rm nl},
\end{equation}
where $V_{\rm ext}$ is the external potential caused by positive
background charges, local parts of the pseudopotential operators and
the potential outside potential barriers. $V_{\rm c}$ is the Coulomb
Hartree interaction part, and $V_{\rm xc}$ is the exchange-correlation
part which we calculate using the local-density approximation
parametrized by Perdew and Zunger \cite{xc3D1,xc3D2}. $V_{\rm bias}$
sets the boundary conditions if a bias voltage is
applied. $\hat{V}_{\rm nl}$ is the nonlocal part of the
pseudopotential operators.

The Hartree potential  is
calculated from the modified Poisson equation 
\begin{equation}
\nabla^2 V_c^i - k_{\rm P}^2 V_c^i = 4 \pi (\rho_+ - \rho) - k_{\rm
P}^2 V_c^{i-1},
\end{equation}
where $k_{\rm P}$ is an adjustable parameter. $k_{\rm P}$ does not
affect the final self-consistent result, but the stability and
convergence of iterations are improved \cite{poisson_k}, because the
Coulomb potential due to charge redistribution between adjacent
iterations is screened.  The non-local pseudopotential is an operator
given by
\begin{equation}\label{pseudo}
\hat{V}_{\rm nl} v({\bf r}) = \sum_{l,m} e_{l,m} \zeta_{l,m}({\bf r})
\int_\Omega \zeta_{l,m}({\bf r}') v({\bf r}') d{\bf r}',
\end{equation}
where $e_{l,m}$ and $\zeta_{l,m}({\bf r})$ are defined using the
Troullier-Martins pseudopotentials \cite{paula-pseudo1,paula-pseudo2}.
Eq. (\ref{pseudo}) uses the projection of the function $v({\bf r})$
(arbitrary function, which in practical calculations is a basis
function) on the atomic-specific function $\zeta_{l,m}$ depending on
the quantum numbers $l$ and $m$ corresponding to the angular momentum.

We have implemented the guaranteed-reduction-Pulay \cite{mix} method
for the mixing of the self-consistent iterations. It uses potentials
from the five previous iterations for computing a new potential in
such a way that the predicted norm of the potential residue is
minimized. The simplest mixing scheme in which potentials are mixed
with a linear feed-back coefficient does not work well in open
systems. The calculations are rather unstable so that quite a small
feed-back coefficient has to be used. This is because the net charge
in the calculation volume $\Omega$ varies during the calculations.

\subsection{Green's function model}

The details of the Green's function method for electron transport in
nanostructures are explained, for example, in Ref.~\cite{gre_xue}. Here we give
only a short introduction to the equations to be solved. The retarded Green's
function $G^r$ is solved from the equation
\begin{equation}\label{greenR}
\big{(}\omega-\hat{H}({\bf r})\big{)} G^r({\bf r},{\bf r}';\omega) =
\delta({\bf r}-{\bf r}'),
\end{equation}
where $\omega$ is the electron energy and $\hat{H}$ is the DFT
Hamiltonian of the system,
\begin{equation}
\hat{H}({\bf r}) = -\frac{1}{2}\nabla^2 + V_{{\rm eff}}({\bf r}).
\end{equation}
When we know $G^r$ we can calculate the so-called lesser Green's
function $G^<$. In the equilibrium when no bias voltage is applied
over the nanostructure it is obtained from
\begin{equation}\label{ele1}
G^<({\bf r},{\bf r}';\omega) = 2f_{L/R}(\omega) \, G^r({\bf r},{\bf
r}';\omega),
\end{equation}
where $f_{L/R}$ are the Fermi functions of the leads. In the
equilibrium, $f_L = f_R$. For a finite bias voltage $f_{L/R}$
we take into account the bias and a more complicated equation for $G^<$
has to be used. To obtain it we write Eq.~(\ref{greenR}) in the form
\begin{equation}\label{eris}
\big{(}\omega-\hat{H}_0-\Sigma_L^r(\omega) - \Sigma_R^r(\omega)
\big{)} \, G^r({\bf r},{\bf r}';\omega)
= \delta({\bf r}-{\bf r}'),
\end{equation}
where $\hat{H}_0$ is the Hamiltonian of the isolated volume $\Omega$
and $\Sigma_{L/R}^r$ are the so-called self-energies of the leads. We
also define
\begin{equation}\label{gammat}
\begin{aligned}
i\Gamma_{L/R} &= \Sigma^r_{L/R} - \Sigma^a_{L/R} = 2i \, {\rm
Im}(\Sigma^r_{L/R}),
\end{aligned}
\end{equation}
and can solve $G^<$ for a finite bias voltage as
\begin{equation}
\begin{aligned}
\label{ele2}
 G^<&({\bf r},{\bf r}';\omega) = \\ & -i f_R(\omega) \int_{\partial
\Omega_R} \int_{\partial \Omega_R} G^r({\bf r},{\bf r}_R;\omega) \,
\Gamma_R({\bf r}_R,{\bf r}'_R;\omega) \\ & \quad \quad \quad \quad
\quad \quad \quad \times G^a({\bf r}'_R,{\bf r}';\omega) \, d{\bf r}_R
\, d{\bf r}'_R \\ &-if_L(\omega) \int_{\partial
\Omega_L}\int_{\partial \Omega_L} G^r({\bf r},{\bf r}_L;\omega) \,
\Gamma_L({\bf r}_L,{\bf r}'_L;\omega) \\ & \quad \quad \quad \quad
\quad \quad \quad \times G^a({\bf r}'_L,{\bf r}';\omega) \, d{\bf r}_L
\, d{\bf r}'_L.
\end{aligned}
\end{equation}
The first and second terms correspond to electrons originating from the
right- and left leads, respectively.  The electron density is
calculated from
\begin{equation}\label{ele_integraali}
\rho({\bf r}) = \frac{-1}{2\pi} \int_{-\infty}^{\infty} {\rm Im}(G^<({\bf
r},{\bf r};\omega)) d\omega
\end{equation}
and the tunneling probability from
\begin{equation}\label{tunneling_probability}
\begin{aligned}
T(\omega) =& \int_{\partial \Omega_L} \int_{\partial \Omega_L}
\int_{\partial \Omega_R}\int_{\partial \Omega_R} 
\Gamma_L({\bf r}_L,{\bf r}'_L;\omega) \,
 G^r({\bf r}'_L,{\bf r}_R;\omega ) \\
\times& \Gamma_R({\bf r}_R,{\bf r}'_R;\omega) \,
G^a({\bf r}'_R,{\bf r}_L;\omega)
\, d{\bf r}_L \, d{\bf r}'_L \, d{\bf r}_R \,d{\bf r}'_R,
\end{aligned}
\end{equation}
from the values of the functions at the boundaries $\partial
\Omega_{L/R}$. Finally the current is determined as
\begin{equation}\label{I}
I = \frac{1}{\pi} \int_{-\infty}^{\infty} T(\omega) \left(
f_{L}(\omega) - f_R(\omega) \right) d\omega.
\end{equation}

We use the FEM in the numerical implementation. Therefore we first
cast Eq.~(\ref{greenR}) in the variational form with open boundary
conditions (for the derivation, see Ref.~\cite{oma}) as
\begin{equation}
  \label{eq:variational_formulation}
  \begin{aligned}
    \int_\Omega &\Big{\{} - \nabla v({\bf r}) \cdot \frac{1}{2} \nabla
    G^r ({\bf r},{\bf r}';\omega) \\ & + v({\bf r}) \big{[}
    \omega-V_{\rm eff}({\bf r}) \big{]} G^r({\bf r},{\bf r}' ; \omega)
    \Big{\}} \,d{\bf r}\\ &- <\hat{\Sigma}_L G^r,v> - <\hat{\Sigma}_R G^r,v>
    \\ =& \, v({\bf r}'),
  \end{aligned}
\end{equation}
where the self energy-operators
\begin{equation}
  \label{eq:dton}
\begin{aligned}
   <\hat{\Sigma}_L G^r,v> = \int_{\partial \Omega_L} \int_{\partial
  \Omega_L} \frac{1}{4} G^r({\bf r}_L', {\bf r}' ; \omega) \\ \times
  \frac{\partial^2 g_e ({\bf r}_L', {\bf r}_L ; \omega)} {\partial
  {\bf n}_L \partial {\bf n}_L'}\, v ({\bf r}_L) \, d{\bf r}_L' d {\bf
  r}_L.
  \end{aligned}
\end{equation}
Above, $g_e$ is the Green's function of the semi-infinite lead in the
domain $\Omega_{L/R}$ with the zero-value condition on the boundary
$\partial \Omega_{L/R}$. In our implementation the leads are described
by a uniform positive background charge and therefore $g_e$ can be
calculated partly analytically. Thus our model means that the leads
are of some kind of ideal generic metals. The important interface
between the nanostructure, e.g. a molecule, and the actual metallic
lead can be described accurately by including some lead metal atoms in
the computational domain $\Omega$. It is also possible to use fully
atomistic leads by calculating numerically $g_e$ for them.

Note that the Eqs.~(\ref{eq:variational_formulation}) and
(\ref{eq:dton}) are analogous to those derivations of the open
boundary conditions in which truncated matrices \cite{gre_datta} are
used. In the continuum limit these two forms give the same
results. However, the weak form is more natural in the FEM formulation
and more suitable for theoretical purposes when analyzing nonlinear
partial differential equations. It is also straightforward to use and
the error control is systematic. Note that this formulation can be
used with any continuous basis set, not only with the FEM. In the
context of basis set methods, the weak formulation case is known as
the Galerkin method.  In practice the Green's functions are
approximated with respect to this basis so that
\begin{equation}
G^r({\bf r},{\bf r}';\omega) \approx
\sum_{i,j=1}^N g_{ij}(\omega) \phi_i({\bf r}) \phi_j({\bf r}').
\end{equation}
The coefficients $g_{ij}(\omega)$ can be solved from
(\ref{eq:variational_formulation}) by choosing $v=\phi_k$ and
evaluating the equations.


\subsection{Finite-element {\it p}-basis}

In the FEM we partition the calculation volume to (in our case,
tetrahedral) sub-domains called elements and the basis functions
$\phi_i$ are constructed using globally continuous (but not
necessarily continuously differentiable) piecewise polynomials with
respect to the finite element mesh. This gives both unique flexibility
of the approximating functions as well as completeness of the basis
with respect to almost any norm. Each basis function has a support that
is concentrated to only a few neighboring elements. This makes the basis
local and results in sparse system matrices.

There are several options of how to choose the finite-element basis
and one has to be careful in achieving acceptable accuracy. The
simplest basis is the linear one. It is easy to implement and works
well, especially in systems with rapidly varying functions.
A typical improvement to this basis is to use node-based
higher-order elements. These elements converge faster to a smooth
solution than the linear ones. However, practically only relatively low
orders, two and three, can be used because of numerical stability
problems.

In this work we have used so-called hierarchical $p$-elements. They
also span higher-order polynomials, but the choice of the local basis
ensures that stability problems do not appear. This
is because the basis functions are chosen so that their derivatives are
close to orthogonal in the $L_2$-norm. The hierarchical nature also
makes it easy to change the order of the basis from element to element
within the same mesh.

The actual FEM implementation consists of a reference element and
reference basis that are mapped separately to each of the elements of
the mesh. Our reference element is a tetrahedron with nodes at the
coordinates $1:(-1,0,0)$, $2:(1,0,0)$, $3:(0,\sqrt{3},0)$ and $4:(0,
\frac{1}{\sqrt{3}}, 2\sqrt{\frac{2}{3}})$. One can easily show that
there exists an affine map taking the reference element to any of the
tetrahedron. The order of our basis is $p$ meaning that in each
element polynomials of the order of $p$ are employed. The basis is
constructed hierarchically. First, there are four linear node basis
functions inside the elements sharing a common node. In the reference
element they are
\begin{equation}
\begin{aligned}
L_1 &= \frac{1}{2}\left(1-\xi
-\frac{\mu}{\sqrt{3}}-\frac{\zeta}{\sqrt{6}}\right)\\ L_2 &=
\frac{1}{2}\left(1+\xi
-\frac{\mu}{\sqrt{3}}-\frac{\zeta}{\sqrt{6}}\right)\\ L_3 &=
\frac{1}{\sqrt{3}}\left(\mu -\frac{\zeta}{\sqrt{8}}\right)\\ L_4 &=
\sqrt{\frac{3}{8}} \zeta\\
\end{aligned}
\end{equation}
where $\xi$ $\mu$ and $\zeta$ are the cartesian coordinates of the
reference element.  Secondly, for $p>$1 we have $6(p-1)$ edge
functions. E.g. for the edge between the nodes 1 and 2
\begin{equation}
\begin{aligned}
N_{i-1}^{(1,2)} = L_1 L_2 \varphi_i(L_2- L_1), \,i=2,...p.
\end{aligned}
\end{equation}
Here one usually sets
\begin{equation}
\begin{aligned}
\varphi_i(\xi) &= \frac{4 \phi_i(\xi)}{1-\xi^2} \\
\phi_i(\xi) &= \sqrt{\frac{1}{2(2i-1)}}(P_i(\xi)-P_{i-2}(\xi))
\end{aligned}
\end{equation}
Above  $P_i$ is the Legendre polynomial of the order of $i$. Third, we
have $2(p-1)(p-2)$ face functions. For example for the face between
the nodes 1,2, and 3 they are
\begin{equation}
\begin{aligned}
N_{i,j}^{(1,2,3)} = L_1 L_2 L_3 P_i(L_2- L_1) P_j(2 L_3- 1),\\
i,j = 0,...,p-3, \,i+j=0,...,p-3.
\end{aligned}
\end{equation}
Fourth, we have $\frac{1}{6}(p-1)(p-2)(p-3)$ bubble functions,
 which are supported only in a single element each. These are
\begin{equation}
\begin{aligned}
N_{i,j,k} = L_1 L_2 L_3 L_4 \, P_i(L_2- L_1) \times \\
P_j(2 L_3- 1) \, P_k(2 L_4-1)\\
i,j,k = 0,...,p-4,\, i+j+k=0,...,p-4.
\end{aligned}
\end{equation}
When $p$-elements are used one must take care of the continuity of the
basis. This is because, for example, the local basis function
$N_{3}^{(1,2)}$ has an orientation on the boundary. The basis includes
the function $\varphi(L_2-L_1)$, not $\varphi(L_1-L_2)$, which would
be another possibility. This means that all the edges in the mesh have
to have information about the direction. Otherwise there is very
likely a continuity problem on some boundaries. In practice, for
tetrahedral elements the orientation problem can be handled for
arbitrary finite-element meshes using only two reference elements
\cite{p_elementit}.

The benefits of selecting the basis described above are rather
clear. The polynomial basis is very easy to realize and has good
approximating properties. For smooth solutions the $p$-basis is known
to give exponential convergence rates with respect to the number of
basis functions used. In the DFT methods the theory is typically
developed to the direction that the solutions are as smooth as
possible. For example, pseudopotential operators are designed so that
they produce as smooth an electron potential as smooth as
possible. This is because the plane-wave basis set needs smooth
solutions in order to work efficiently.  On the other hand, in the
case of non-smooth solutions one can benefit from the piecewise nature
of the FEM basis allowing one to approximate even singular solutions
to some extent.  Moreover, the finite element mesh can be refined in
regions where solution changes rapidly.  When modeling molecules there
is also a lot of empty space in the calculation domain. It is then
practical to use large elements in the empty space and smaller ones
near atoms.


\subsection{Linear algebra methods}

The use of the Green's function method is computationally demanding in
comparison to explicit wave-function methods.  Since the main
computational burden of our method is to find a subset of the
coefficients of the Green's function in question, a special
consideration must be taken when choosing the methods of linear
algebra to be used. The eigenvalue problems in explicit wave-function
methods are typically solved by iterative methods. In our case it is
better to use direct solvers, because a set of linear equations needs
to be solved.  We have opted for the frontal method widely used in the
solution of sparse linear systems \cite{duff-reid, duff} and extremely
suitable for finite-element matrices. The actual implementation is
ME47 of the Harwell Subroutine Library (HSL) \cite{hsl} (see
Refs.~\cite{watson,umfpack,superlu} for other similar approaches). In
the frontal method, one first finds a permutation of the sparse matrix
aiming to minimize the fill-in resulting from the factorization
process. Next, a LU-decomposition (or Cholesky-decomposition for
symmetric problems) of the matrix $A$ is found, and finally, two
systems with triangular coefficient matrices, $Lz=b$ and $Ux=z$ (where
$U=L^T$ for symmetric problems) are solved. To find all the required
coefficients of the solution we must vary the right-hand side $b$ of
the system.

For a three-dimensional problem the size of the linear system can grow
so large that the CPU-time and memory requirements of different
systems have to be addressed. The main question is how large systems
can be calculated using these methods so that the calculation time for
a single self-consistent iteration is not too large. Currently a
system of several tens of thousands of unknowns can be solved in a
commodity-CPU cluster environment.

In detail, the Green's function method includes a computation of the
elements for the inverse of a sparse matrix, so that the calculation
time requirements increase relatively fast with the system size. A
classical complexity result for the solution (and inversion) of a
general $N \times N$ system with a direct method is $O(N^3)$. However,
for sparse systems and modern frontal methods this bound is too
pessimistic \cite{george-liu}. The CPU time requirement depends on the
fill-in of the inversion problem.
For very simple cases one can show that the key statistic of the
problem, the number of non-zeros ({\it nnz}) present in the factors
$L$ and $U$, satisfies $nnz(L) \sim nnz(U) = O(N \log{(N)})$
\cite{george-liu}.
Then the solution of each of the systems requires $O(nnz(L) + nnz(R))$
floating-point operations, and in the worst case we must solve these
with $N$ different right-hand sides effectively giving us the inverse
of the matrix $A$, so that the total cost is $O(N(nnz(L) + nnz(R)))$.
However, in modern computer systems the complexity is
not the only relevant measure since the performance may be highly
nonlinear (see, e.g. \cite{atlas, goto} for an example on
BLAS-tuning).

Another topic related directly to the performance of modern computer
systems is the relation between processor power and memory
bandwidth. This is especially true for the computation of the Green's
function where the actual bottleneck is the lack of available
memory bandwidth in commodity-based cluster systems used in
calculations, not the floating-point performance of the processor itself.  

It is likely that a better performance can be achieved by upgrading
several parts of the algorithms. First, the current parallel solver is
implemented using the Message Passing Interface (MPI)
\cite{mpi}. However, in Symmetric multiprocessor (SMP) systems it is
likely that well-designed OpenMP \cite{openmp} (or similar)
parallelism would reduce the need for data transfer and thus increase
performance. It would also decrease the memory requirements of the
problems. Second, at the moment the solution of the Green's function
is computed varying one vector on the right-hand side at a time. A
better performance could be obtained if the equations could be solved
for multiple right-hand sides at a time allowing the use of BLAS3
routines. Finally, it is likely that computations would benefit from a
computer system having a larger memory bandwidth than our present
commodity-based one.


\section{Example systems}\label{examples}

\subsection{Atomic wire}

Using the atomic force microscope or the mechanically controlled break
junction technique, a chain of atoms can be made of certain metals
\cite{stm_lanka}.  It has been observed that the conductances of atom
chains vary as a function of the number of atoms in the chain
\cite{atomilanka}. The conductances of these systems have been studied
also theoretically in several works.  In order to benchmark our
results against other calculations, we use Na-atom chains as test
systems. They have been simulated in several previous studies
\cite{sim, lyhyt, biaksen_kanssa, lee} using different
models. According to these calculations the conductances of the wires
show even-odd oscillation as a function of the number of atoms in
the wire.

\begin{figure}[hbt]
\begin{center}
\epsfig{file=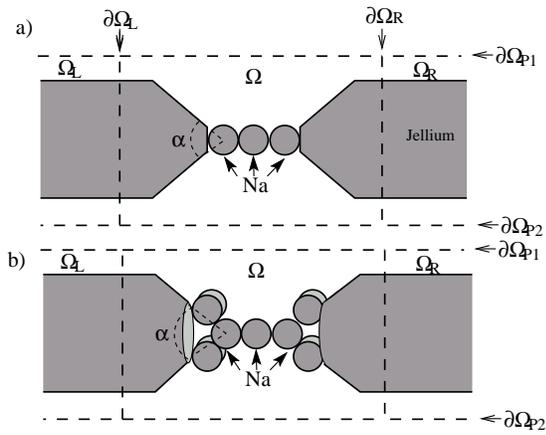,width=0.4\textwidth}
\end{center}
\caption{\label{Na_ketjut} Two models of the Na-atom chain. a) In model
{\bf A}, Na atoms are directly connected to the cone-shape leads. b) In
model {\bf B}, there are four Na atoms  as  squares at the tips of the
leads.}
\end{figure}

In our setup, the Na-atom chain  is located between two leads, with
the lead shape defined by a 70$^\circ$ cone angle (see
Fig.~\ref{Na_ketjut}). We consider two different connections of the
atom chain to the electrodes. In model {\bf A} we have just three Na
atoms between the jellium leads. This resembles closely the system
used in Ref.~\cite{lyhyt}. In model {\bf B}, there are four
Na atoms at the tips of the leads in a square form.  This makes the
connection between the atom chain and the leads more realistic. This
kind of structure is modeled also in Ref.~\cite{biaksen_kanssa}.

\begin{figure}[hbt]
\begin{center}
\epsfig{file=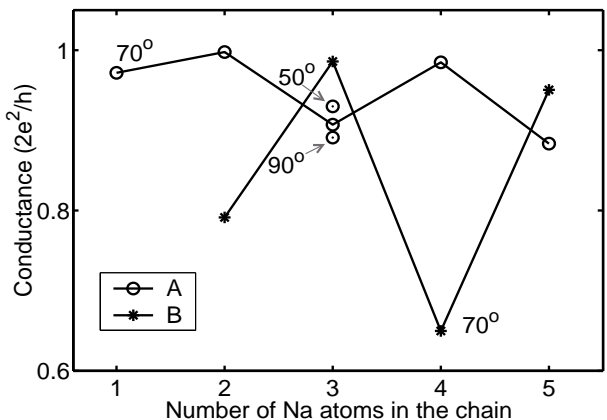,width=0.45\textwidth}
\end{center}
\caption{\label{ketju_kon} Conductance through the Na-atom chain as
  function of the number of Na atoms in the chain. The results for
  systems {\bf A} and {\bf B} (see Fig.~\ref{Na_ketjut}) are denoted
  by circles and stars, respectively. For system {\bf A} with three
  Na atoms, results corresponding to 50$^\circ$ and 90$^\circ$ cone
  angles $\alpha$ are also given.}
\end{figure}

\begin{figure}[hbt]
\begin{center}
\epsfig{file=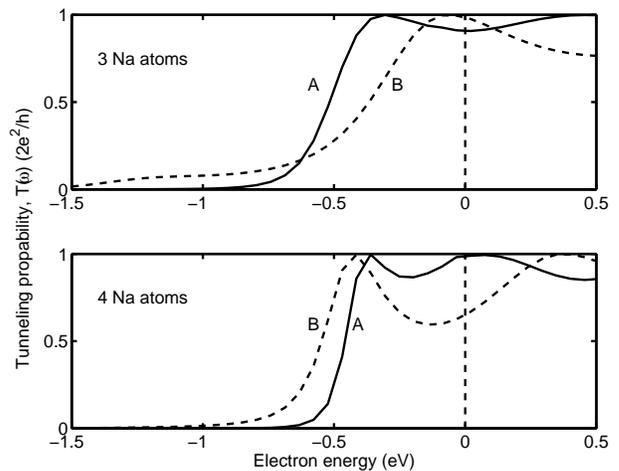,width=0.45\textwidth}
\end{center}
\caption{\label{ketju_tun} Tunneling probability from
  Eq.~(\ref{tunneling_probability}) for three- and four- atom chains
  between two semi-infinite jellium leads. The solid and dashed lines
  correspond to systems {\bf A} and {\bf B} shown in
  Fig.~\ref{Na_ketjut}, respectively. The Fermi-level is marked by dashed
  vertical lines. The cone angle $\alpha$=70$^\circ$}
\end{figure}

The conductances as a function of the number of chain atoms for
systems {\bf A} and {\bf B} are shown in Fig.~\ref{ketju_kon}. In the
Na-atom chain, electrons have only one conducting mode so that the
conductance can be one conductance quantum $2e^2/h$ at maximum.  Both
systems {\bf A} and {\bf B} exhibit conductance oscillations as a
function of the number of atoms. These oscillations arise from
resonance states in the atom chain. Depending on the position of the
resonances relative to Fermi-level the conductance has either a
maximum or minimum value, so that the maxima and minima correspond to
approximately half and fully occupied resonance states,
respectively. The oscillation is within the range of 0.9 - 1.0
$\times 2e^2/h$ for system {\bf A} and 0.6 - 1.0 $\times 2e^2/h$ for
system {\bf B}. The difference between the oscillation amplitudes is
due to different strengths of the connection of the chains to the
leads.  System {\bf B} has a weaker coupling to the leads than system
{\bf A}. Weak connections make the resonances also sharper, as is seen
in the tunneling probability in Fig.~\ref{ketju_tun}.  In contrast to
Ref. \cite{oma_johto}, we do not see a strong lead-shape
dependence in the conductance. The widening of the cone angle lowers
the conductance as the edges of the wire become sharper.

The electron tunneling probabilities through chains of three- and
four- atom systems {\bf A} and {\bf B} are shown in
Fig.~\ref{ketju_tun}. The probability function $T(\omega)$ is defined
in Eq.~(\ref{tunneling_probability}). The conductance of the system in
the zero-bias limit can be read at the Fermi-level. Here, as well as
in Fig~\ref{ketju_kon}, we see that the conductance oscillations for
systems {\bf A} and {\bf B} are in a different phase. This is because
in system {\bf B} the atom chain is effectively shorter than in {\bf
A}, as the first and the last chain atom are partly inside the square
of the four Na atoms.

When we compare the conductance oscillations of system {\bf A} (see
Fig.~\ref{ketju_kon}) to those in Ref.~\cite{lyhyt} obtained by using
semi-infinite jellium leads with planar surfaces ($\alpha =
180^\circ$), we see that the even-odd oscillations in the conductance
are in the same phase. In the case of system {\bf B} we can directly
compare the tunneling probability of Fig.~\ref{ketju_tun} with those
in Ref.~\cite{biaksen_kanssa} where the atom chain is connected also
through a square of four Na atoms to jellium. The phase and the
amplitude of the conductance oscillations of these results are in 
good agreement with our values in Fig.~\ref{ketju_kon}.  Now
that we have satisfied ourselves that the method provides a good model
for electron transport we can consider a more interesting and
demanding example.


\subsection{Thin insulating layer}

The general increase in the performance of microelectronic devices in
the past few decades has been made possible by continuous transistor
scaling - based on a reduction in the thickness of the gate dielectric
in typical metal-oxide-semiconductor field-effect transistors
(MOSFET). At present the process has reached a bottleneck, as further
reduction leads to a large increase in leakage current due to direct
tunneling across the thin silicon dioxide (SiO$_2$) layer. Several
possible approaches to resolve this are being considered
\cite{highk_nature}, but retaining conventional MOSFET design remains
an economically attractive choice, and a leading option is just to
replace SiO$_2$ with another oxide of higher dielectric constant
(high-$k$). A high-$k$ oxide would provide higher effective
capacitance to a comparable SiO$_2$ layer, hence allowing thicker
layers to be used to reduce losses due to tunneling. The specific
choice of oxide is determined by a set of requirements
\cite{huff03_reqs} based on both the intrinsic properties of the grown
oxide and its integration into the fabrication process, and at present
hafnium oxide (HfO$_2$) remains a leading candidate.

In order to study to transport properties of thin HfO$_2$ films we
have simulated the growth of the oxide on a silicon surface via first
principles molecular dynamics \cite{hakala05_int}. Here we consider
three model interfaces: (i) a nonstoichiometric oxide interface ({\bf
C}), which is basically metallic due to Hf-Hf and Hf-Si bonds across
the interface; (ii) a stoichiometric oxide interface ( {\bf D}), which
has a localized state in the band gap due to a few Hf-Hf bonds; (iii)
a more idealistic interface ({\bf E}), which remains insulating if no
defects are present. The last model is based on the interface used in
Ref. \cite{fonseca03_int}, but slightly reduced in size to make it
computationally manageable \cite{gavartin05_int}. These models were
calculated with periodic boundary conditions with k-points on the
boundaries $\partial \Omega_{P1/P2/P3/P4}$. The effective potentials
have been calculated for systems {\bf C} and {\bf D} using the gamma
point, and for system {\bf E}, four $k$-points. All the tunneling
probabilities $T(\omega)$ are calculated using four $k$-points, which
was enough to converge the probabilities to a good accuracy.

As shown in Fig. \ref{nano_malli}, the interface models are positioned
between two leads. The charge density in the leads is chosen so that
in the right lead $r_s = 2$ (electron density $n_e = 3/(4 \pi r_s^3)$),
representing a metal, and in the left one $r_s = 3.1$, representing
doped-silicon - as in a standard MOSFET design.

\begin{figure}[htb]
\begin{center}
\epsfig{file=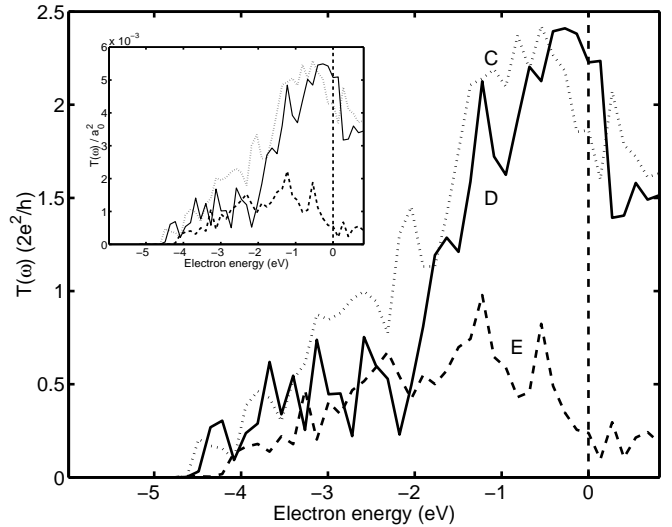,width=0.49\textwidth}
\end{center}
\caption{\label{tun_good_bad} Tunneling probability $T(\omega)$
  through thin HfO$_2$ layers. Results for system {\bf C} (dotted
  line) - nonstoichiometric interface, system {\bf D} (solid line) -
  stoichiometric interface and system {\bf E} - ideal interface are
  shown. The inset shows $T(\omega)$ normalized with the conducting
  area enabling the comparison of actual insulating properties of
  different systems.}
\end{figure}

The tunneling profiles of the systems are shown in
Fig.~\ref{tun_good_bad}. Here it is seen that systems {\bf C} and {\bf
D} show clearly metallic behavior, with a large tunneling probability
at the Fermi energy. Although in principle, the stoichiometric
interface ({\bf D}) has a much lower density of metallic bonds, it is
clear that both in interfaces {\bf C} and {\bf D} around two channels
dominate the transport. The localized defect state in the band gap of
system {\bf D} plays an equivalent role in transport to the metallic
bonds in interface {\bf C}.

As expected, the tunneling probability for the more ideal interface
{\bf E} is an order of magnitude smaller at the Fermi energy than
those for interfaces {\bf C} and {\bf D}. Yet we also see that it
remains significant - this is largely due to the structure of the
interface itself \cite{fonseca03_int}. Although bulk HfO$_2$ is a
wide-bandgap insulator, at the interface it exists as almost
tetragonal HfSiO$_4$, and the effective band gap is actually smaller
than that of bulk silicon below the interface. This means that there
is a negative conduction band offset between silicon and HfO$_2$, and
no real barrier for leakage. Although some of this is caused by the
underestimation of the band gap in the DFT, this also reduces the
silicon band gap (although the effect is not systematic).

\begin{figure}[htb]
\begin{center}
\epsfig{file=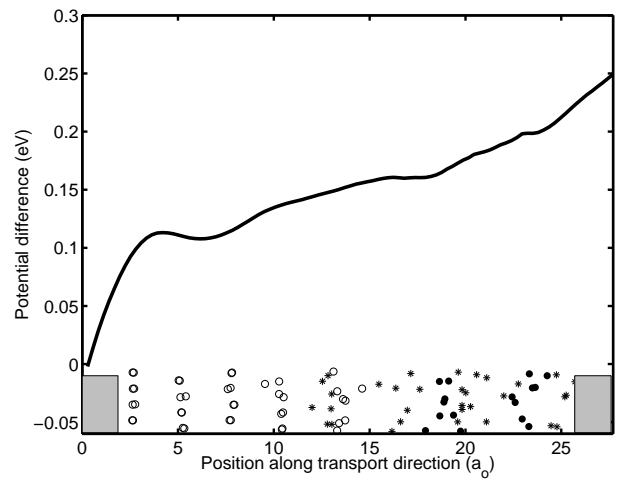,width=0.45\textwidth}
\end{center}
\caption{\label{pot_diff} Change of the average effective potential in
  interface {\bf D} when a 0.25 V bias voltage is applied over the
  HfO$_2$ layer.  In every position along the transport direction the
  effective potential is averaged over the perpendicular-coordinates.
  Atom positions are indicated: open circles are silicon, filled
  circles, and stars hafnium and oxygen atoms, respectively. The gray
  areas mark the positions of the leads.}
\end{figure}

The poor performance of interface {\bf D} can also be seen in its
capacity for dropping the potential.  Fig. \ref{pot_diff} shows the
potential change for 0.25 V applied bias voltage. The potential drop
across HfO$_2$ is less than 0.05 eV, demonstrating that the oxide
hardly perturbs the electron flow from the right lead. The potential
drops fastest at the right hand side of HfO$_2$-layer where pure
HfO$_2$ exists, and much more slowly in the thin layer of SiO$_2$
formed due to diffusion of oxygen. The large drop at the lead and
silicon atoms is just an artifact of the boundary conditions of the
Coulomb part of the effective potential.

\begin{figure}[htb]
\begin{center}
\epsfig{file=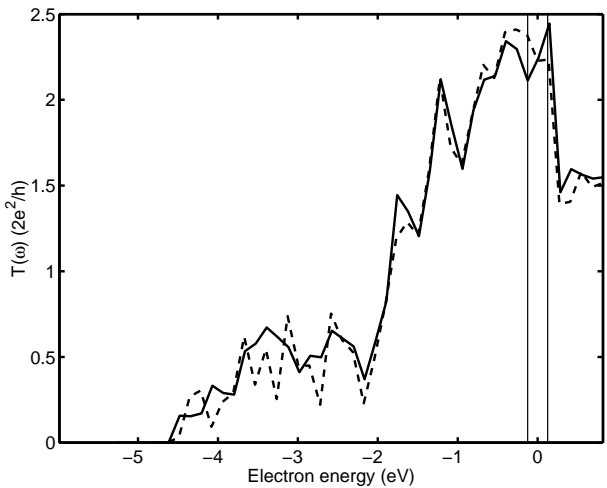,width=0.45\textwidth}
\end{center}
\caption{\label{tun_bias} Tunneling probability through a thin HfO$_2$
  layer.  Results for interface {\bf D} with 0.25 V bias (solid line)
  and zero bias (broken line) voltages shown. The vertical lines show
  the positions of the Fermi levels in the left and right leads for
  the biased system.  Between them is located the ``so-called''
  conductance window [see equation~(\ref{I})]. The Fermi level of the
  non-biased system is in the middle of these lines.}
\end{figure}

In the rigid band approximation (used for example in
Ref.~\cite{fonseca03_int}) it is assumed that the shape of the
tunneling probability stays constant and is only shifted in energy so
that $T(\omega,V_{\rm bias}) = T(\omega+\eta V_{\rm bias})$, where
$\eta$ is the ratio of potential drop at the other end of the
nanostructure to the total drop over the nanostructure. In
Fig.~\ref{tun_bias} we have studied how well this approximation works
for interface {\bf D}. The curves are plotted so that the zero-bias
Fermi level is in the middle of the left and right Fermi levels of the
biased interface. This corresponds to the symmetric case with $\eta=
0.5$. 
We see that the tunneling probability curves roughly coincide. This
indicates that potential drops symmetrically over the nanostructure
and the rigid-band approximation gives a rather reasonable result.


The above results show, in agreement with previous calculations
\cite{fonseca03_int} that tunneling through a more ideal, insulating
interface is still significant due to a negative band offset with
silicon. Since the only HfO$_2$ interfaces providing significant band
offsets to silicon were built very idealistically (i.e. assuming no
significant atom migration nor interfacial SiO$_2$ growth)
\cite{fiorentini02_int,peacock04_int}, this indicates that fabricating
a \emph{good} interface directly between silicon and HfO$_2$ is very
difficult. A more viable alternative maybe to sacrifice somewhat in
dielectric constant, and grow HfO$_2$ onto a pre-existing SiO$_2$
layer. These possibilities will be explored in more detail in a
further work \cite{hakala05_int}.

\section{Conclusions}\label{discussion}

In this paper we present a finite-element implementation of the
non-equilibrium Green's function method which is combined to the
density-functional theory. Although the Green's function method is
computationally demanding, we demonstrate that by using hierarchical
$p$-elements, large, physically relevant systems become
tractable. More importantly, our method offers a much more rigorous
control of accuracy than is usually possible in transport
calculations.

We demonstrate the functionality of our implementation with two kinds
of systems, the sodium atom chain wire and the silicon-HfO$_2$
interface. For the atom chain, we show that the method reproduces the
previous results of other Green's function transport methods. This
gives us confidence to apply it to the  more complex system: a thin layer
of hafnium oxide on a silicon substrate. Here we show that the transport
properties are an even more sensitive indicator of the role of defects
than the electronic structure. Comparison of stoichiometric and
non-stoichiometric HfO$_2$ oxide layers demonstrates that even one or
two defects in a stoichiometric interface can result in tunneling
comparable to that of a fully metallic non-stoichiometric interface.

\acknowledgments

We are grateful to J. L. Gavartin and L. R. C. Fonseca for providing
us with access to their interface structures, and for helpful
discussions. We acknowledge the generous computer resources from the
Center for Scientific Computing, Espoo, Finland.  This research has
been supported by the Academy of Finland through its Centers of
Excellence Program (2000-2005). We have used the Harwell Subroutine
Library in our calculations.




\bibliography{hafnia,zirconia,tech,paula}
\bibstyle{apsrev}


\end{document}